# End-effects in rapidly rotating cylindrical Taylor-Couette flow

Rainer Hollerbach* and Alexandre Fournier†

*Department of Mathematics, University of Glasgow, Glasgow, G12 8QW, United Kingdom
†Laboratoire de Géophysique Interne et Tectonophysique, Université Joseph Fourier
BP 53, 38041 Grenoble Cedex 9, France

**Abstract.** We present numerical simulations of the flow in a rapidly rotating cylindrical annulus. We show that at the rotation rates relevant to the magneto-rotational instability, the flow is strongly constrained by the Taylor-Proudman theorem. As a result, it is controlled almost entirely by the end-plates. We then consider two possible options for minimizing these end-effects, namely (i) simply taking a very long cylinder, and (ii) splitting the end-plates into a series of differentially rotating rings. Regarding option (i), we show that the cylinder would have to be hundreds of times as long as it is wide before end-effects become unimportant in the interior. Since this is clearly not feasible, we turn to option (ii), and show that in order to obtain a smooth angular velocity profile, the end-plates would have to be split into around ten rings. If the end-plates are split into fewer rings, perhaps 3–5, the angular velocity profile will not be smooth, but will instead consist of a series of Stewartson layers at the boundaries from one ring to the next. We suggest therefore that the instabilities one obtains in this system will be the familiar Kelvin-Helmholtz instabilities of these Stewartson layers, rather than the magneto-rotational instability. At best, one might hope to obtain the MRI superimposed on these Kelvin-Helmholtz modes. Any subsequent interpretation of results is thus likely to be quite complicated.

## INTRODUCTION

The flow between concentric rotating cylinders, the so-called Taylor-Couette problem, is one of the most widely studied systems in classical fluid dynamics [1]. The magnetic extension of this problem, in which the fluid is taken to be electrically conducting and a magnetic field is applied along the cylinders, is equally fundamental in classical magnetohydrodynamics [2]. This magnetic problem has received particular attention recently as a prototype – and possible experimental realization [3–5] – of the magneto-rotational instability (MRI) that is the topic of this book. In this contribution we wish to focus attention on the role of the end-plates that would necessarily be present in any real experiment. We will show that in the parameter regime relevant to the MRI these end-effects have an enormous influence throughout the entire fluid. We question therefore whether the MRI could actually be observed in this system.

Consider two concentric cylinders of radii $r_i$ and $r_o$, rotating at rates $\Omega_i$ and $\Omega_o$ respectively (which we will both take to be positive, that is, the cylinders are rotating in the same direction). If the cylinders are infinitely long – very easy to assume theoretically, but rather more difficult to build experimentally! – one possible solution for the flow in

the gap between them is given by

$$\mathbf{U} = (ar + b/r)\hat{\mathbf{e}}_\phi, \tag{1}$$

where

$$a = \frac{\Omega_o r_o^2 - \Omega_i r_i^2}{r_o^2 - r_i^2}, \qquad b = \frac{r_i^2 r_o^2}{r_o^2 - r_i^2}(\Omega_i - \Omega_o). \tag{2}$$

The essence of the classical Taylor-Couette problem then is to consider the ways in which this basic state can yield to different solutions, such as Taylor vortices, wavy or spiral Taylor vortices, and many other possibilities. It is this rich variety of solutions that makes this system so fascinating.

In contrast, in the MRI problem we start with this same basic state – which is still a valid solution even in the presence of a magnetic field along the $z$-axis – but now we are interested in very different instabilities of it. In particular, we would like to choose the various parameters such that the solution (1) is hydrodynamically stable, but magnetohydrodynamically unstable. According to the Rayleigh criterion, (1) will be stable provided $a > 0$, that is, $\Omega_i r_i^2 < \Omega_o r_o^2$. However, according to MRI theory [6], in the presence of a magnetic field (1) may nevertheless be unstable, provided only that $\Omega_i > \Omega_o$. In principle therefore it would seem almost trivially straightforward to observe the MRI in the laboratory: simply rotate your cylinders somewhere in the intermediate range $1 < \Omega_i/\Omega_o < r_o^2/r_i^2$, and then adjust the strength of the magnetic field. If the field is too weak (or too strong) the solution (1) will be stable, but if the field strength is just right (1) will be unstable, and one has obtained the MRI.

Of course, in reality the situation isn't quite so simple (otherwise someone would most likely already have stumbled upon this phenomenon purely by accident). The complication that makes such an MRI experiment so difficult is that not only must the relative rotation rates of the two cylinders lie in this range $1 < \Omega_i/\Omega_o < r_o^2/r_i^2$, but beyond that their absolute magnitudes must also be enormous. According to the WKB analysis of Ji *et al.* [3, 4], and confirmed by the full solution of Rüdiger *et al.* [7, 8], the rotation rates must be so large that the Reynolds number $Re = (r_o - r_i)^2 \Omega_o / \nu$ exceeds $10^6$ or even $10^7$. More precisely, the relevant quantity is the magnetic Reynolds number $Rm = (r_o - r_i)^2 \Omega_o / \eta = Re \cdot (\nu/\eta)$, which must be $O(10)$ before the MRI can set in. It is the extremely small magnetic Prandtl numbers $Pm = \nu/\eta$ of sodium ($10^{-5}$) or gallium ($10^{-6}$) that then lead to these large values of $Re$.

In view of this extremely rapid rotation of both cylinders, we believe it is useful to analyze this problem in a way that may be unfamiliar from the MRI perspective, but is standard in geophysical fluid dynamics [9], in which a rapid overall rotation is typically a dominant component. Scaling length by the gap width $(r_o - r_i)$, time by the inverse rotation rate $\Omega_o^{-1}$, and $\mathbf{U}$ by $(\Omega_o - \Omega_i)(r_o - r_i)$, the Navier-Stokes equation in the frame co-rotating with the outer cylinder becomes

$$\frac{\partial}{\partial t}\mathbf{U} + Ro\,\mathbf{U}\cdot\nabla\mathbf{U} + 2\hat{\mathbf{e}}_z \times \mathbf{U} = -\nabla p + E\nabla^2 \mathbf{U}, \tag{3}$$

where the Ekman and Rossby numbers

$$E = \frac{\nu}{\Omega_o(r_o - r_i)^2} \quad \text{and} \quad Ro = \frac{\Omega_i - \Omega_o}{\Omega_o} \quad (4)$$

measure the overall and differential rotation rates, respectively. Comparing with the parameters introduced above, we find then that $E = Re^{-1}$, so obtaining the MRI would require Ekman numbers as small as $10^{-7}$. For $Ro$ we similarly find that the range we are interested in is $0 < Ro < (r_o^2 - r_i^2)/r_i^2$.

The main advantage of formulating the problem in this way is that it allows us to separate the effects of the overall rotation, as measured by $E$, and the differential rotation, measured by $Ro$. In particular, we will find that the structure of the basic state is controlled almost entirely by $E$, with $Ro$ playing a relatively minor role. In the next two sections we therefore set $Ro = 0$, corresponding to an infinitesimal differential rotation, and consider the influence of $E$ alone. Because (3) is then linear, we are able to reduce $E$ down to $10^{-6}$. We will show that for these extremely small values it is almost impossible to achieve the desired profile (1); the entire flow is controlled by end-effects, which impose a very different profile. In the final two sections we then consider the influence of finite $Ro$, and also the implications of these results for these planned MRI experiments.

## RIGID END-PLATES

As an example, let's consider a cylinder having $r_i = 1$, $r_o = 2$, and height $h = 5$, with the end-plates co-rotating with the outer cylinder. Remembering that (3) has been formulated in this same reference frame rotating at the rate $\Omega_o$, and that **U** has been nondimensionalised using the differential rotation, we recognize that the appropriate boundary conditions are

$$U_\phi = 1 \quad \text{at} \quad r = r_i, \qquad U_\phi = 0 \quad \text{at} \quad r = r_o \quad \text{and at} \quad z = \pm h/2, \quad (5)$$

and of course zero everywhere for $U_z$ and $U_r$. Fig. 1 shows the solutions for $Ro = 0$ (as already noted above) and $E = 10^{-4}$, $10^{-5}$ and $10^{-6}$, computed using the spectral element code of Fournier *et al.* [10].

Turning to the angular velocity first, we see that it does *not* consist of anything like the smooth profile (1) we were hoping for. Instead, it is zero almost everywhere in the interior, with all of the adjustment occurring in a narrow shear layer right at the inner cylinder. The reason for this peculiar structure is the well-known Taylor-Proudman theorem, stating that in rapidly rotating systems the flow will tend to align itself along the axis of rotation. More formally, if we take the curl of (3) and use $\partial_t$, $Ro$, $E \ll 1$ (in fact $\partial_t$ and $Ro$ are identically zero here), we obtain $\frac{\partial}{\partial z}\mathbf{U} \approx 0$. With this result, this behavior seen in Fig. 1 follows quite naturally; having $\omega = 0$ in the interior will then satisfy both the Taylor-Proudman theorem as well as the boundary conditions on the end-plates.

Of course, right at the inner cylinder the solution cannot be $\omega = 0$, since the boundary condition there is $\omega = 1$. The structure of this shear layer within which the solution adjusts to this inner boundary condition was deduced by Stewartson [11], and shown to scale as $E^{1/4}$. The results in Fig. 1 are consistent with this scaling, with the layer three times thinner at $10^{-6}$ than at $10^{-4}$.

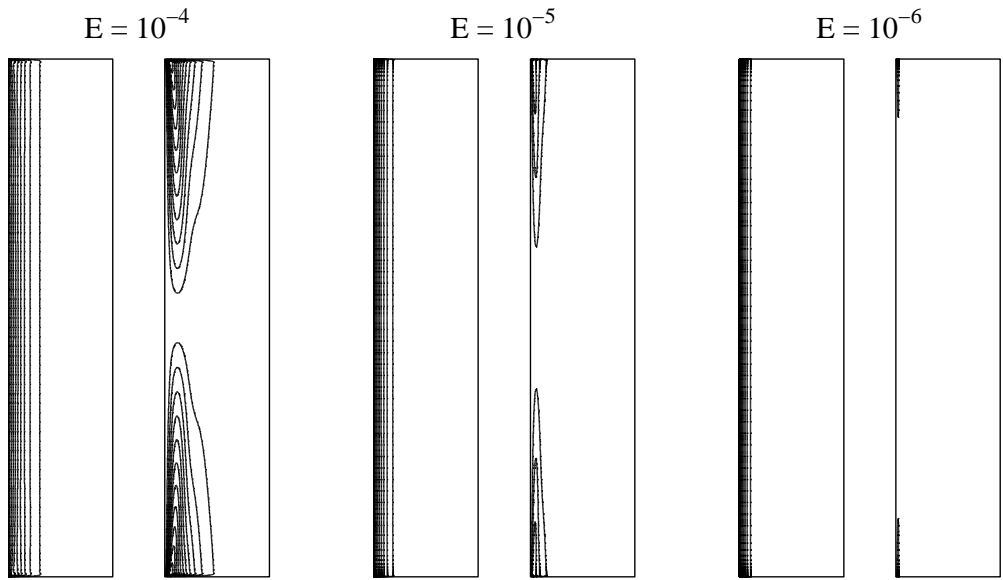

**FIGURE 1.** Within each pair, the left panel shows contours of the angular velocity $\omega$, with a contour interval of 0.1, and the right panel shows the streamfunction of the meridional circulation $\Psi$, with a contour interval of $3 \cdot 10^{-4}$. As the Ekman number is reduced, we see how the entire structure is increasingly concentrated toward the inner boundary, in perfect agreement with the $E^{1/4}$ and $E^{1/3}$ scalings derived by Stewartson. In addition, we see how the meridional circulation becomes weaker and weaker, again consistent with the $E^{1/2}$ asymptotic scaling. Finally, we note that both $\omega$ and $\Psi$ have so-called Ekman layers at the end-plates. Because these layers scale as $E^{1/2}$, they are so thin as to be almost invisible in these plots. Right at the end-plates we really do have the proper $\omega = \Psi = 0$ boundary conditions though.

Finally, Fig. 1 also shows the associated meridional circulation, consisting of an inward flow at the end-plates, and a return flow in the interior. This flow too is concentrated near the inner cylinder. Its detailed structure is slightly different from that for $\omega$, including also an $E^{1/3}$ scaling. The amplitude of this entire circulation decreases with $E$ though, scaling as $E^{1/2}$. This is therefore a relatively unimportant part of the flow, and will not be considered further.

Summarizing Fig. 1 then, we see that if $E = 10^{-4}$ were sufficiently small to obtain the MRI, our cylinder here might already be good enough, with this Stewartson layer sufficiently broad that it might still be close enough to the smoother profile we really wanted. If we need to reduce $E$ down to $10^{-6}$ or even $10^{-7}$ though, it is clear that this cylinder will not do. So, suppose we try taking a longer cylinder, in the hope of minimizing the end-effects. Fig. 2 presents profiles of $\omega$ as a function of $r$, in the midplane $z = 0$, for $h = 5$ (the solutions in Fig. 1), 10 and 20. We see that increasing $h$ helps somewhat, but not nearly enough, particularly at the smaller values of $E$. Of course, this was only to be expected; it is known [9] that the influence of the Taylor-Proudman theorem extends a distance $O(E^{-1/2})$ into the interior. In order for the end-plates not to dominate everything, one would therefore need to take $h$ to be several thousand!

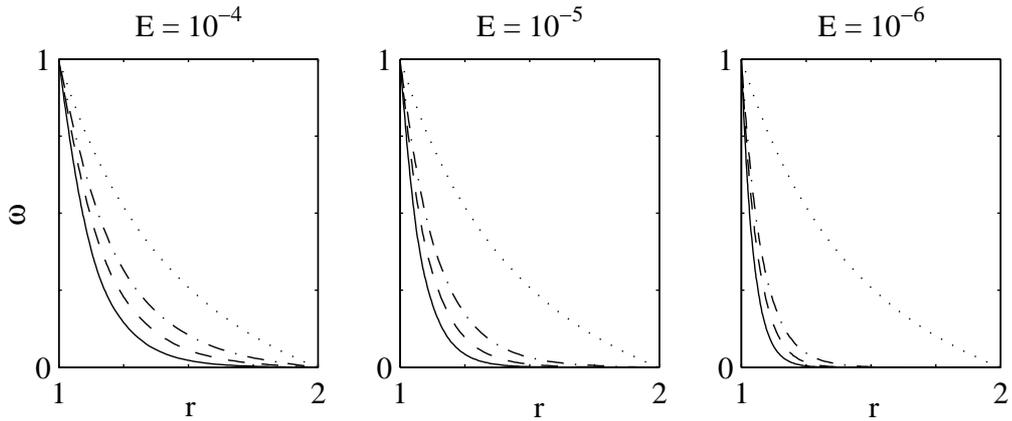

**FIGURE 2.** The angular velocity in the midplane $z = 0$. In each panel the solid lines are for $h = 5$, the dashed lines for $h = 10$, and the dash-dotted lines for $h = 20$. The dotted lines indicate the profile (1) one would obtain for an infinitely long cylinder. Extrapolating the $h = 5$, 10 and 20 results, we see therefore that for $E = 10^{-4}$ one would need $h \approx 100$ to achieve the desired profile, but for $E = 10^{-6}$ one would need $h \approx 1000$, in agreement with the expected asymptotic scaling.

Incidentally, note also that this is very different from the classical Taylor-Couette problem, where end-effects play a comparatively minor role [12, 13]. The difference is that there the outer cylinder is typically at rest, so the Taylor-Proudman theorem does not apply. If the entire system is rotating as rapidly as $E = 10^{-6}$ though, the Taylor-Proudman theorem applies with a vengeance, so end-effects extend far deeper into the interior, and end up controlling everything, as we saw in Figs. 1 and 2.

## SPLIT-RING END-PLATES

Taking a cylinder whose length is several thousand times its diameter is clearly not feasible. Following Kageyama *et al.* [5], let us therefore consider the possibility of splitting the end-plates into several distinct, independently rotating rings. For example, rather than $\omega = 0$ on the end-plates, suppose we impose instead the boundary conditions

$$\omega = \left\{ \begin{array}{ll} 1 & \text{for} \quad 1 \leq r < 1.25 \\ 2/3 & \text{for} \quad 1.25 \leq r < 1.50 \\ 1/3 & \text{for} \quad 1.50 \leq r < 1.75 \\ 0 & \text{for} \quad 1.75 \leq r \leq 2 \end{array} \right\}. \qquad (6)$$

That is, instead of having all of the adjustment from 1 inside to 0 outside occurring right at $r = 1$, we have split it up, with 1/3 each occurring at $r = 1.25$, 1.5 and 1.75. If this works at all in yielding a smooth profile in the interior, one could then obviously adjust the precise values to obtain any desired profile, including (1).

Fig. 3 shows the results. We see that at $E = 10^{-4}$ we do indeed have a smooth profile, and even at $10^{-5}$ it is still reasonably smooth. However, once we reduce $E$ down to $10^{-6}$, we are tending toward a step profile, with three separate Stewartson layers at $r = 1.25$, 1.5 and 1.75. The $E = 10^{-4}$ profile is smooth only because the Stewartson layers are still so thick in that case that three of them essentially fill the whole gap, so they do not yet manifest themselves as distinct layers. At $E = 10^{-6}$ they are so thin though that one would need to split the end-plates into perhaps ten rings to obtain a comparably smooth profile.

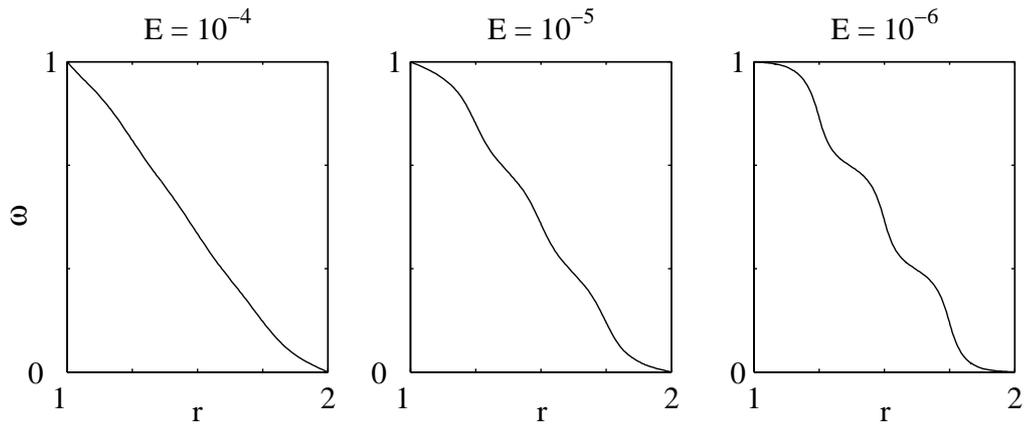

**FIGURE 3.** As in Fig. 2, profiles of $\omega$ in the midplane $z = 0$, but now with the boundary conditions (6) imposed on the end-plates.

These results also explain why Kageyama *et al.* did not recognize the existence of these Stewartson layers. Because they focussed on getting the relative differential rotation right (*Ro* in our notation here) their overall rotation was still several orders of magnitude less than it ought to be, so their Stewartson layers were like our $E = 10^{-4}$ layers, sufficiently thick that three or four of them filled the entire gap, and didn't form distinct layers at all. As our results here show though, as soon as one reduces $E$ into the appropriate regime, these layers do manifest themselves quite clearly even for split-ring end-plates.

## FINITE ROSSBY NUMBER

All the results up to now have been for $Ro = 0$, corresponding once again to an infinitesimal differential rotation. We remember though that really we are interested in $O(1)$ Rossby numbers. And at first glance one might suppose that the results would be very different in that regime. In particular, if $Ro = O(1)$ it is not clear whether the Taylor-Proudman theorem still applies, and if it does not, then all of the preceding analysis would appear to be irrelevant.

In fact, with a little thought we can soon convince ourselves that the Taylor-Proudman theorem should still apply, at least for Rossby numbers only slightly beyond zero. The argument is as follows: If $\mathbf{U}$ is independent of $\phi$ (trivially true for an axisymmetric basic state) and almost independent of $z$ (as we saw above), then so is $\mathbf{U} \cdot \nabla \mathbf{U}$. If $\mathbf{U} \cdot \nabla \mathbf{U}$ depends only on $r$ though, it can be balanced by a pressure-gradient (just as it is in the desired profile (1)). That is, the inclusion of the inertial term $Ro\,\mathbf{U} \cdot \nabla \mathbf{U}$ in (3) should have relatively little effect on the solutions. That this is indeed the case is seen in the sample calculation in Fig. 4. We see therefore that the Taylor-Proudman theorem continues to apply, and correspondingly these Stewartson layers continue to develop (although at finite $Ro$ we are no more able to achieve $E = 10^{-6}$ than Kageyama et al. were, and hence also cannot simulate the desired parameter regime directly).

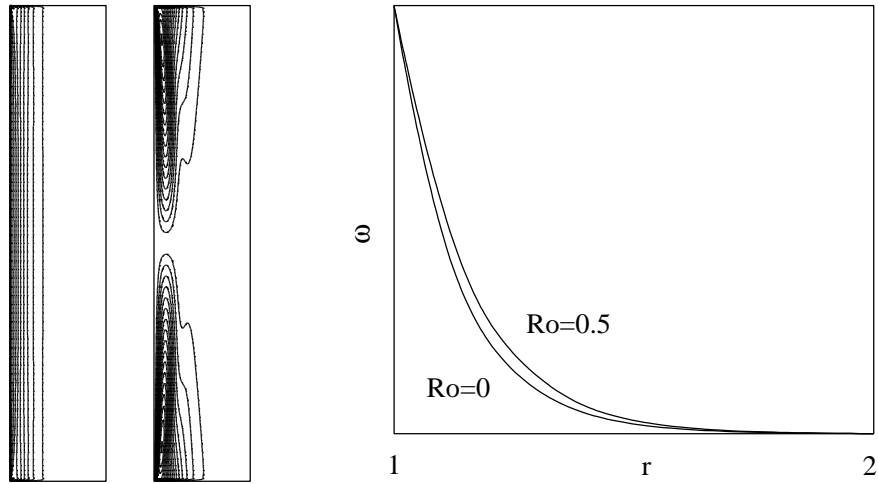

**FIGURE 4.** The solution at $h = 5$, $E = 10^{-4}$ and $Ro = 0.5$. The first two panels show $\omega$ and $\Psi$. Comparing with Fig. 1, we see that the meridional circulation has been strengthened somewhat, but is still much weaker than the angular velocity, which is hardly changed at all. More quantitatively, the final panel again shows profiles of $\omega$ in the midplane, and shows that increasing $Ro$ from 0 to 0.5 does broaden the Stewartson layer somewhat, but not nearly enough.

For sufficiently large $Ro$ solutions do actually exist that break the Taylor-Proudman theorem. For example, Kageyama et al. present flows having a very strong meridional circulation that is both $z$- and even time-dependent. There are two difficulties with these solutions. First, while we may have broken the Taylor-Proudman theorem, and hence avoided these undesirable Stewartson layers, these solutions are also not quite what we really wanted. In particular, the desired profile (1) is actually independent of $z$. So breaking the Taylor-Proudman theorem may have eliminated the Stewartson layers, but at the cost of introducing an equally undesirable $z$-dependence.

Second, and more importantly, while these large $Ro$ solutions may exist in the sense that one can compute them with an axisymmetric code such as ours here or that of Kageyama et al., they would not exist in a real MRI experiment. Other, non-axisymmetric instabilities would set in long before the critical Rossby numbers for these solutions are reached. In particular, these Stewartson layers are susceptible to a Kelvin-Helmholtz instability, in which the initially circular shear layer adopts a wavy structure instead. There is a considerable body of literature already on these instabilities of Stewartson layers, including experimental [14–16], analytical [17, 18] and numerical [18, 19] work. All of these agree that the critical Rossby number for the onset of these Kelvin-Helmholtz modes scales roughly as $E^{3/4}$. At $E = 10^{-6}$ the critical Rossby number would therefore be around $10^{-4.5}$, which is long before any of these modes set in that break the Taylor-Proudman theorem (which these Kelvin-Helmholtz modes do not).

# CONCLUSION

With all these various pieces in place, we believe we can now predict what will happen in the parameter regime we originally set out to explore, namely $E \leq 10^{-6}$ and $Ro = O(1)$. In particular, let's imagine fixing $E$ and gradually increasing $Ro$. For sufficiently small $Ro$ these Kelvin-Helmholtz instabilities will not yet have set in, and the solution will consist simply of a series of Stewartson layers, one at every split in the rings making up the end-plates. And as we saw in section 3, splitting the end-plates into enough rings that these layers merge into a smooth profile is unfortunately next to impossible. We will therefore have a step profile, separated by a sequence of distinct shear layers. As we further increase $Ro$ then, we will rather quickly reach this critical value $Ro_c \propto E^{3/4}$ where these layers go unstable, and adopt non-axisymmetric, wavy structures.

Extrapolating finally from $Ro = O(E^{3/4})$ to $O(1)$ is of course quite a leap, but $Ro$ is then so many orders of magnitude supercritical that it seems highly likely that these wavy structures will give way to turbulent shear layers. The one positive aspect of all this is that in the process the underlying shear will almost certainly be considerably reduced. For example, Hollerbach et al. [20] found that Rossby numbers only 2 or 3 times supercritical were already enough to significantly broaden the Stewartson layer again. It is quite conceivable therefore that the average profile will actually end up looking rather like the profile (1) we wanted all along. The flow will be turbulent rather than laminar though, so any subsequent identification of the MRI is likely to be considerably more difficult and ambiguous than originally hoped for.